# Machine Learning aided k-t SENSE for fast reconstruction of highly accelerated PCMR data

Grzegorz Tomasz Kowalik[1], Javier Montalt-Tordera[1], Jennifer Steeden[1], Vivek Muthurangu[1].

[1]UCL Institute of Cardiovascular Science, Centre for Cardiovascular Imaging, 30 Guildford Street, London, UK WC1N 1EH

**Corresponding author:**

Dr Grzegorz Tomasz Kowalik, PhD

Centre for Cardiovascular Imaging

UCL Institute of Cardiovascular Science

London, UK

email: grzegorz.kowalik.09@ucl.ac.uk

**Author contributions**

<u>Kowalik GT:</u> conceptualization, methodology, data curation, formal analysis, investigation, resources, software, validation, visualization, writing - original draft
<u>Montalt-Tordera J:</u> development of machine learning framework software
<u>Steeden J.</u> writing - review & editing, resources
<u>Muthurangu V:</u> funding acquisition, writing - review & editing, conceptualisation, resources

**Running title:**

ML aided k-t SENSE

# Abstract

Purpose: We implemented the Machine Learning (ML) aided k-t SENSE reconstruction to enable high resolution quantitative real-time phase contrast MR (PCMR). Our approach uses a U-net to create a high quality prior for the k-t SENSE reconstruction. We additionally tested a new version of U-net (U-net M) for improved restoration of magnitude images corrupted due to the very high data undersampling.

Methods: A residual U-net and our U-net M (that additionally uses the time averaged data as the spatial signal distribution map) were used to generate the high resolution x-f space estimate for k-t SENSE regularisation prior. The networks were trained on 816 retrospectively gated PCMR magnitudes and were judged on their ability to generalise to real undersampled data. The in-vivo validation was done on 20 real-time 18x prospectively undersmapled $GAS_{perturbed}$ PCMR data. The ML aided k-t SENSE reconstruction results were compared against the free-breathing Cartesian retrospectively gated sequence and the compressed sensing (CS) reconstruction of the same data.

Results: The standard U-net results were inadequate for k-t SENSE priors with significant distortions to imaged signal. Consequently, only U-net M was used in the in-vivo study.

In general, the ML aided k-t SENSE generated flow curves that were visually sharper than those produced using CS. In two exceptional cases, U-net M predictions exhibited blurring which propagated to the extracted velocity curves. However, there were no statistical differences in the measured peak velocities and stroke volumes between the tested methods.

The ML aided k-t SENSE was estimated to be ~3.6x faster in processing than CS.

Conclusion: The ML aided k-t SENSE reconstruction enables artefact suppression on a par with CS with no significant differences in quantitative measures. The timing results suggest the on-line implementation could deliver a substantial increase in clinical throughput.

**Introduction**

Phase Contrast Magnetic Resonance (PCMR) is a proven method of measuring blood flow in the clinical environment [1, 2]. Although most PCMR acquisitions are cardiac gated, real-time methods [3] have become more common for rapid free-breathing assessment of blood flow [3-5]. However, acquisition of real-time PCMR data requires significant data undersampling and several methods have been used to remove the resultant aliasing. One of the most powerful methods is Compressed Sensing (CS) [6], which has been used for high resolution real-time PCMR [7]. Unfortunately, CS reconstructions are computationally intensive with long reconstruction times and this limits clinical utility.

A recent development has been the use of machine learning (ML) for MR image reconstruction [8]. In the simplest form, it can be formulated as an image restoration problem. For example, aliasing artefacts can be suppressed through ML based de-noising of MR images [9], as long as these artefacts are incoherent in nature. Incoherent aliasing and subsequent image restoration can be achieved through golden ratio imaging and convolutional neural networks (CNN), i.e. U-net [10]. This method has been used to reconstruct undersampled golden angle radial real-time data [11]. The benefit of these methods over CS, is that the computational load is shifted to the training stage rather than at the time of image reconstruction. This yields substantial increase in the reconstruction speed as compared to CS [12, 13]. However, these methods are not immediately applicable to the complex representation of PCMR data and neglect the intrinsic redundancy of multi-coil data [13].

More advanced uses of machine learning for MR image reconstruction [14] combine a deep learning approach with the conventional inverse problem optimization. In these methods, CNNs are used to learn and enforce data driven regularization of models that incorporate the data consistency term and are trained end-to-end. Although very promising, these approaches require raw k space data for training, which is rarely available in large amounts.

We propose a hybrid reconstruction framework combining the formulation of parallel imaging as a regularized inverse problem (k-t SENSE), with CNN derived priors. The k-t SENSE algorithm uses both parallel imaging and spatio-temporal correlations (prior knowledge about spatial distribution of temporal frequencies in the signal, in x-f space) for reconstruction. The algorithm's performance heavily relies on how well an x-f estimate matches with an imaged signal. In conventional k-t SENSE these priors are derived from low spatial, high temporal resolution fully sampled training data. These priors are readily available for radial trajectories (by extracting the fully sampled centre of k space) but are more difficult to acquire with Cartesian and spiral sampling. Spiral trajectories are especially attractive to PCMR for their short TE values and effitient k space sampling. This restricts k-t SENSE's utility for high resolution PCMR acquisitions. In our framework, we adopt a de-noising CNN model to create a high spatio-temporal resolution signal magnitude estimate from the alias corrupted magnitude data. These high resolution estimates are then used as the prior for the k-t SENSE algorithm to produce the complex-valued results.

ML aided k-t SENSE has the potential to match the CS reconstruction quality, in a fraction of the processing time. In this study, we test the ML aided k-t SENSE reconstruction on data acquired with an 18x accelerated Golden Angle Spiral Perturbed ($GAS_{perturbed}$) real-time PCMR sequence. The perturbed spiral trajectories produce highly incoherent aliases that have noise like characteristics. This sampling pattern has been shown [7] to be well suited to CS reconstruction, enabling free-breathing blood flow quantification with high spatio-temporal resolution. It is also well suited to the de-noising task. However, the conventional U-net magnitude restoration has not been tested at such high acceleration factors and it may not provide adequate artefact suppression. Therefore, we propose a modified version that further utilizes information about the spatial distribution of the signal.

The aims of the study are: i) to compare the ability of both conventional and modified U-nets to produce high quality imaging priors, ii) perform clinical validation of the ML aided k-t SENSE for quantitative PCMR, and iii) compare the results and reconstruction times with CS.

## 1. Material and methods

The k-t SENSE reconstruction utilising parallel imaging for data on arbitrary trajectories [15] is formulated as the regularised optimisation problem:

$$arg \min_{\rho_{x,f}} \frac{1}{2} \left\| E_{x,f \to k,t} \rho_{x,f} - y_{k,t} \right\|_2^2 + \lambda \left\| M_{x,f}^{-2} \rho_{x,f} \right\|_2^2 \tag{1}$$

where $y_{k,t}$ is the acquired multi-frame ($t$ - time) under-sampled multi-coil k space data. $\rho_{x,f}$ is the spatial- ($x$) temporal frequency ($f$) signal that is being searched for, that minimises the sum of $\ell_2$ norms ($\|\cdot\|_2^2$). The first argument is the data fidelity term that ensures consistency between $\rho_{x,f}$ and $y_{k,t}$. $E_{x,f \to k,t}$ is the imaging system matrix encapsulating the inverse Fourier Transform along the time domain ($\mathcal{F}_{t \to f}^{-1}$), combination with coil sensitivity estimates ($S$), Fourier Transform into k space ($\mathcal{F}_{x \to k}$) and sampling onto the acquisition trajectory positions. The second norm $\left\| M_{x,f}^{-2} \rho_{x,f} \right\|_2^2$ represents Tikonov regularisation with $M_{x,f}^{-2}$ being the inverse of the estimated signal intensity on a diagonal. Due to under-sampling the system is ill-posed. However, this regularisation gives preference to solutions similar to the estimate $M_{x,f}$ with the regularisation parameter ($\lambda$) controlling the trade of between the two norms.

In this work we propose the following estimate of the signal intensities:

$$M_{x,f}^2 = \left| \mathcal{F}_{t \to f} U_w \left( \left| S^H \mathcal{F}_{x \to k}^{-1} y_{k,t} \right| \right) \right|^2 \tag{2}$$

Here, $U_w(\cdot)$ is a CNN trained with $w$ weights to perform de-aliasing [11] on the magnitudes of the gridded k space data ($\left| S^H \mathcal{F}_{x \to k}^{-1} y_{k,t} \right|$).

### 1.1. Neural Network design

In this study we implemented two versions of the CNN architecture known as U-net [10]. The first version is a modified residual U-net [11] (shortened to $U_w$) that is comprised of three multi-scale decomposition levels, furnished with a pair of encoding and decoding convolutional units, and a skip connection between them. Each convolution unit contained two 3D convolutional layers with 3x3x3 filter size, with the rectified linear unit (ReLU) as the activation function. Max-pooling was used to down-sample the resolution, while 3D transpose convolution with stride 2x2x2 and no activation function was used to up-sample the resolution. The result of the last decoding unit was reduced to a single channel output with a 3D convolutional layer (1x1x1 filter size, no activation function) and added to the input as the residual update.

The second version called U-net M (shortened to $U_w^M$) expands $U_w$ to make full use of the available information from time-resolved MRI (Fig. 1). Specifically, an acquisition with constantly rotating complementary trajectories (i.e. GAS$_{perturbed}$) enables creation of an artefact free image by performing time-averaging of this data. In this version, the difference between the magnitude of the time averaged data and each gridded frame was used as a second input channel. In $U_w^M$ the max-pooling operation was replaced by applying stride 2x2x2 in the last 3D convolutional layer of the encoding units. Lastly, the result of the last convolutional layer was added to the magnitude time average (the second input) as the temporal signal update.

The CNN architectures were implemented using TensorFlow 2.0 [16] in Python 3. All predictions were run off-line. Due to memory limitations the input size was fixed to patches of 56x56x56.

### 1.2. Training and data preparation for ML

The training data consisted of magnitude images from 816 retrospectively Cardiac gated uniform spiral PCMR data sets acquired in the aortic view [17]. The data were collected with varying spatial: 1.7 ± 0.1 x 1.7 ± 0.1 mm and temporal: 15.2 ± 10.8 ms resolution. The average heart rate was: 75 ± 16 beats per minute.

The following data curation and augmentation steps were performed in order to create the real-time ground truth data for training of the CNNs. The images were median filtered (size: 3x3x3) to reduce noise. The filtered retrospectively cardiac gated data were interpolated to match the real-time in-vivo assessment parameters (matrix: 256x256, pixel size: 1.76x1.76 mm, 96 frames of ~26.6 ms). A continuous acquisition was emulated by cycling through the r-r interval. Additionally, the number of training sets was augmented by applying in-plane rotations (-45°, 0° and 45°). The preparation of the training data was implemented in C++ to reduce processing time.

To create the paired undersampled, artifact-contaminated training data, the ground truth images were sampled onto the target GAS$_{perturbed}$ trajectory described in [7]. The synthetic k space data was then re-gridded to create the alias corrupted data sets. For $U_w^M$, the second input was created by taking the time average of the corrupted complex data sets. The magnitude values were extracted from all the generated data volumes. These were normalised to the [0, 1] range and separated into overlapping patches of 56x56x56 pixels (resulting in 50 evenly spaced patches). 80% of the final 122400 sets (three rotations and 50 patches, for each of the 816 sets) were randomly selected and used for the CNNs training with the rest constituting the validation set. A batch size of 32 was selected and ADAM optimiser was run for 18 epochs with the mean average error (MAE) loss function. Additionally during the training, Structural Similarity Index Measure (SSIM) was calculated between the CNNs' results and the truth to assess the training's performance.

The CNN selection criterion was the capability of the trained model to generalise to prospectively undersampled in-vivo PCMR data. As there is no reference available, the CNN results for all of the prospective PCMR data underwent a visual qualitative assessment (GTK, over 10 years of expertise) for sufficient artefact suppression and lack of signal distortions.

### 1.3. In-vivo study

The in-vivo data acquired in the previous study [7] consisting of 20 pediatric patients referred for cardiac clinical MR (7 females and 13 males; age range: 6 - 16 years, median: 12.5 years) were used. The National Research Ethics Committee approved the study (Ref: 06/Q0508/124) and a written consent was obtained from all patients or legal guardians of children. The imaging data consisted of reference standard free-breathing Cartesian retrospectively gated PCMR data (FOV: 350x262 mm, voxel: 1.82x1.82x6.0 mm, TR/TE: 4.4/1.9 ms, Flip Angle: 30°, VENC: 200 cm/s, averages: 2, GRAPPA: 2, temporal resolution: 18.5 ms) and the real-time $GAS_{perturbed}$ PCMR data (FOV: 450x450 mm, voxel: 1.76x1.76x6.0 mm, Flip Angle: 20°, VENC: 200 cm/s, TR/TE: 6.7/1.9 ms, temporal resolution: ~26.6 ms, total of 270 frames). The real-time $GAS_{perturbed}$ PCMR data was reconstructed using the CS recossntruction [7], as well as the ML aided k-t SENSE technique. All data was collected in the ascending aorta just above the sino-tubular junction.

### 1.4. ML aided k-t SENSE

The $GAS_{perturbed}$ real-time PCMR data acquired in the previous study were reconstructed with the new two stage ML aided k-t SENSE (Fig. 2). The 270 PCMR frames (~7.2 s) were reconstructed in three blocks of 96 (including 6 overlapping) frames. The blocks were combined applying averaging to the overlapping frames.

The first stage is estimation of the signal magnitude ($M_{x,f}^2$) using the trained CNN. Both flow encoded ($y'_{k,t}$) and compensated ($y''_{k,t}$) data blocks were processed as described in Eq. (2). The time averaged compensated data was used to estimate coil sensitivity maps [18] that are necessary to perform coil weighted combination of the individual under-sampled multi-coil gridded frames ($\rho'_{x,t}$, $\rho''_{x,t}$). The magnitude of the combined time average and corrupted frames were normalised to [0, 1] range and separated into 56x56x56 patches for processing by a CNN. The patched predictions were then recombined in to full size frames applying averaging to the overlaps. To improve the resulting SNR, the predicted encoded and compensated magnitudes were temporally sorted and low-pass filtered by applying the Tukey filter in the temporal frequency space (cut-off frequency: ~44%). The central half of the filtered x-f space was extracted and used as the signal intensity estimate ($M_{x,f}^{-2}$) for the second stage of the reconstruction.

In the second stage of ML aided k-t SENSE the linear conjugate gradient solver was used to solve the minimisation problem Eq. (1) and produce the final encoded ($\rho'_{x,t}$) and compensated ($\rho''_{x,t}$) data results from the undersampled k space data ($y'_{k,t}$, $y''_{k,t}$). The regularisation level and the number of iterations for the k-t SENSE reconstruction were set to $\lambda$ = 0.01 and five respectively. These were based on a visual assessment of imaging results from a single subject.

The gridding operation in the first stage of the new reconstruction was implemented in C++ utilising NVIDIA CUDA to speed up the processing. The coil combination, extraction of magnitudes and generation of x-f estimates were implemented in Julia [19]. The k-t SENSE was implemented as the extension to the previously reported SENSE reconstruction for real-time flow quantification [5]. The NVIDIA K40 GPU card was utilised in the processing.

### 1.5. Flow quantification

The aorta was segmented on the magnitude images using a semi-automatic method based on the optical flow registration [20] with manual operator correction using in-house plugins for Horos software (Horos, free LGPL license at Horosproject.org,

sponsored by Nimble Co LLC d/b/a Purview in Annapolis, MD USA). The resultant regions of interest (ROI) were transferred to the phase images to produce flow and velocity curves. Maximum velocity was taken as the peak of the velocity curve. Stroke volume (SV) was calculated by integrating the resultant flow curve over a single r-r interval. As multiple heartbeats are evaluated with real-time PCMR, SV and peak velocity are averaged across all complete r-r intervals.

### 1.6. Image quality

Quantitative image quality was assessed by estimating SNR, VNR and edge sharpness (ES). All quantitative analyses were carried out by using in-house plug-ins for Horus software. True quantification of SNR and VNR in images acquired with non-Cartesian trajectories is nontrivial owing to the uneven distribution of noise [21, 22]. Following the previous study, estimated SNR and VNR were calculated as described [23]. In summary, a region of interest (ROI) was drawn in stationary tissue, and estimated noise was calculated as the average standard deviation of the pixel intensity ($\sigma_s$) or velocity ($\sigma_v$) through time. Final estimates of SNR and VNR were made by dividing the mean signal intensity from a ROI drawn at peak systole by $\sigma_s$ and $\sigma_v$, respectively.

ES was calculated in peak systole by measuring the average maximum gradient of the normalised pixel intensities across the aortic wall. The image data was resampled onto evenly spaced perpendicular lines crossing the vessel border (marked with the ROIs used to extract the velocity data). Lanczos resampling [24, 25] was used with 0.5 mm step between samples on the lines on a distance of 20 mm. Further, 'the smooth noise robust differentiation' [26] was applied to extract the maximum gradient on the projections.

For the real-time data the SNR, VNR and ES measurements were performed in all peak systole frames and the averaged values were used for comparisons.

### 1.7. Statistical Analysis

All statistical analysis was performed using R software (R Foundation for Statistical Computing, Vienna, Austria) and a p-value of less than 0.05 indicated a significant difference. All the results are expressed as mean ± standard deviation. Differences between the three imaging techniques were assessed using the one-way repeated measures analysis of variance (ANOVA). The imaging techniques were treated as the repeated measures factor. Significant results were further investigated with post-hoc pairwise comparison using the Tukey method.

## 2. Results

### 2.1. CNN model training

The $U_w$ achieved a MSE of 24.6e-3 and a SSIM of 0.91 for the validation set, while $U_w^M$ achieved a slightly higher MSE of 26.9e-3 and the same SSIM of 0.91.

However, it was observed that $U_w$ failed to recover the underlying signal when applied to the prospectively undersampled in-vivo PCMR test set. The visual assessment uncovered signal distortions in all of the 20 cases. These included visually distinguishable boundaries between the adjacent patches and smaller or larger parts of the signal were missing in all cases. In contrast, these boundaries and signal removal issues were not present in the $U_w^M$ results.

In two prospective cases, the $U_w$ completely removed heart structures (Fig. 3). This demonstrates that the $U_w$ model was unable to reliably provide x-f maps. Thus, the final trained $U_w^M$ was used for the first stage of the ML aided k-t SENSE reconstruction as earlier described.

### 2.2. Feasibility of ML aided k-t SENSE

The in-vivo PCMR data was reconstructed off-line with the new ML aided k-t SENSE technique and the previously reported CS reconstruction in order to compare the reconstruction times. Timings for the preparation of the gridding matrices and accessing data from a disk were neglected. This was justified by the fact the timings would be the same for both techniques.

For each block of 90 frames the CS reconstruction required ~59 s., whilst the ML aided k-t SENSE reconstruction required ~16.6 s. This included ~3.9 s to grid the k space samples, ~9.9 s for neural network's predictions and ~1s for the x-f signal estimate generation. The linear conjugate gradient for k-t SENSE required ~1.9s .

### 2.3. In-vivo flow quantification

Examples of velocity and flow curves for the reference breath-hold Cartesian PCMR data and the GAS$_{perturbed}$ data reconstructed with both CS and ML aided k-t SENSE techniques are shown in Fig. 4. The curves extracted from the ML aided k-t SENSE results were visually sharper and exhibited less flattening of the peaks and troughs.

However, there were two case in which $U_w^M$ provided a temporally blurred estimate of x-f space. This projected into k-t SENSE and resulted in blurring of the curves.

There were no statistical differences (p > 0.2) in peak velocity measured between the Cartesian reference (72.4 ± 18.0 cm/s), the real-time data with CS reconstruction (72.3 ± 18.6 cm/s) and the real-time data with ML aided k-t SENSE reconstruction (73.2 ± 18.3 cm/s). The ML aided k-t SENSE technique had an insignificant positive bias for the peak velocity measures compared to the Cartesian reference (bias: 0.8 cm/s, limits: -4.8 to 6.4 cm/s, Fig. 5) and compared to the CS reconstruction (bias: 1.0 cm/s, limits: -3.9 to 5.9 cm/s, Fig. 5). As previously reported CS had no significant difference in the peak velocity measure compared to the Cartesian reference, with slightly narrower limits of agreement (bias: -0.1, limits: -4.4 to 4.1 cm/s, Fig. 5).

Aortic stroke volume measurements showed no statistical difference between the ML aided k-t SENSE data (72.0 ± 24.1 ml) and both the Cartesian reference (73.2 ± 23.7 ml, p = 0.4) and the CS reconstruction (71.4 ± 23.4 ml, p = 0.8) results. The ML aided k-t SENSE technique had slightly better agreement with the Cartesian reference, but broader limits of agreement than the CS technique (bias: -1.2 ml, limits: -11.4 to 9.1 ml, vs. bias: -1.8 ml, limits: -9.4 to 5.8 ml for CS, Fig. 5).

### 2.4. In-vivo image quality

Representative imaging results are shown in Fig. 6 and Supporting Information Video S1-6. There was a significant (p < 1e-5) difference in estimated SNR between the reference Cartesian (110.3 ± 38.6) and both the ML aided k-t SENSE (29.5 ± 18.0) and the CS (52.7 ± 25.8) data. In addition, ML aided k-t SENSE had significantly lower SNR than CS reconstruction (p < 0.05). Estimated VNR results showed no significant difference (p > 0.3) between the ML aided k-t SENSE (14.1 ± 5.3) and both Cartesian (16.1 ± 7.6) and CS (16.3 ± 5.8) results.

ES was highest in the Cartesian images (0.14 ± 0.03 mm$^{-1}$). There was no significant (p > 0.3) difference between the CS (0.12 ± 0.04 mm$^{-1}$) and the ML aided k-t SENSE images (0.11 ± 0.03 mm$^{-1}$). The ES results for the ML aided k-t SENSE data were significantly (p = 0.01) lower than the Cartesian data. There was no statistical difference (p > 0.2) between the CS and the Cartesian ES measures.

## 3. Discussion

In this work, we addressed the issue of clinically acceptable processing times for highly accelerated real-time PCMR data. Previously, a CS reconstruction [7] has been shown to deliver high quality results for accurate blood flow quantification. However, the reconstruction times, even using GPU processing, were longer than 1 min. As a solution, we proposed a combination of Machine Learning and an iterative image reconstruction to provide accurate flow quantification with clinically acceptable reconstruction times.

The following factors played a role in the development of our hybrid solution. A purely CNN approach, as previously used for real-time cine image restoration [11], would require adoption to complex valued flow data. More sophisticated end-to-end trained deep recursive architectures that alternate between a CNN and data consistency term could be used to reconstruct flow data. Such approaches have been shown to outperform CS [14], but require raw k space data that is rarely available in the clinical setting. We present a middle ground solution that does not require raw multi-coil training data but retains data consistency with measurements. Specifically, we use a CNN to supress artefacts in the magnitude images of highly undersampled real-time PCMR data. The CNN generated results are then used as a regularisation prior to aid the well-established parallel imaging reconstruction, k-t SENSE. Our approach falls under the deep prior learning for inverse problem solving [8, 9, 27]. This can be classified as a simpler version of the model base deep learning [28].

Previous studies have used U-nets to suppress artefacts in radial magnitude images with moderate acceleration. This study uses very heavily undersampled GAS$_{perturbed}$ sampling with much greater magnitude image corruption. A standard U-net based magnitude restoration may prove insufficient in this case. For this reason, we tested a conventional U-net ($U_w$) and a modified version ($U_w^M$) designed to better leverage the data. There was little difference in the performance of $U_w$ and $U_w^M$ on synthetic undersampled data. However, the networks generalised very differently to real-life data. With prospectively undersampled PCMR data, we observed $U_w$ removes parts of the signal and in two cases removed the heart completely. These signal aberrations rendered the standard $U_w$ model inadequate for our problem. The $U_w^M$ leverages the time averaged data as a spatial signal distribution map. This greatly improved image reconstruction of the prospective data and prevented the removal of heart structures seen with $U_w$. For this reason, we used $U_w^M$ to generate regularisation priors for k-t SENSE.

Due to the high acceleration factor, there was a drop in quantitative image quality (lower SNR and ES) in the ML aided k-t SENSE images as compared with the CS results of the same data. The $\ell_1$ regularised sparse signal recovery algorithms (i.e. CS) are well known to provide superior noise attenuation compared to $\ell_2$ based inverse problem solutions (i.e. k-t SENSE) that can amplify noise. However, there were no statistically significant differences in VNR, peak mean velocity and stroke volume measures from the Cartesian, CS and the ML aided k-t SENSE results. Thus, our approach can provide clinically robust real-time flow quantification. The major benefit is the ~3.6x faster than CS reconstruction, which is highly conducive to clinical translation.

### 3.1. Limitations

Our work does suggest that models trained on coil combined magnitude data may have difficulty in generalising to undersampled multi-coil data. Further improvements may be possible by using complex valued CNNs [29]. However, most MR assessments are image based with phase information being stripped away before storing. This can make collection of training data for complex CNNs challenging.

### 3.2. Conclusion

We have presented the initial validation of a ML aided k-t SENSE reconstruction for clinically relevant real-time PCMR flow quantification. In the work we address the long reconstruction times associated with CS reconstruction of high spatio-temporal real-time PCMR data. The presented $U_w^M$ model enables sufficient estimation of x-f space coefficients for use as a regularisation term for k-t SENSE. ML aided k-t SENSE enabled artefact suppression on a par with CS with no significant differences in the quantitative blood flow measures. The timing results suggest the on-line implementation could deliver a substantial increase in clinical throughput.

**15****Acknowledgements**

This work was supported in part by British Heart Foundation grant: NH/18/1/33511. JAS and JMT are funded under the UKRI Future Leaders Fellowship (MR/S032290/1). GTK and JMT are also part funded by Heart Research UK (RG2661/17/20).

**FIGURES**

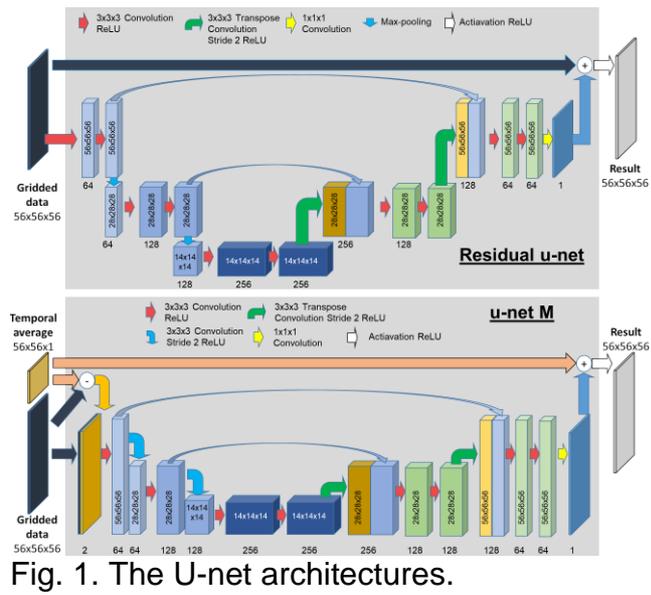

Fig. 1. The U-net architectures.

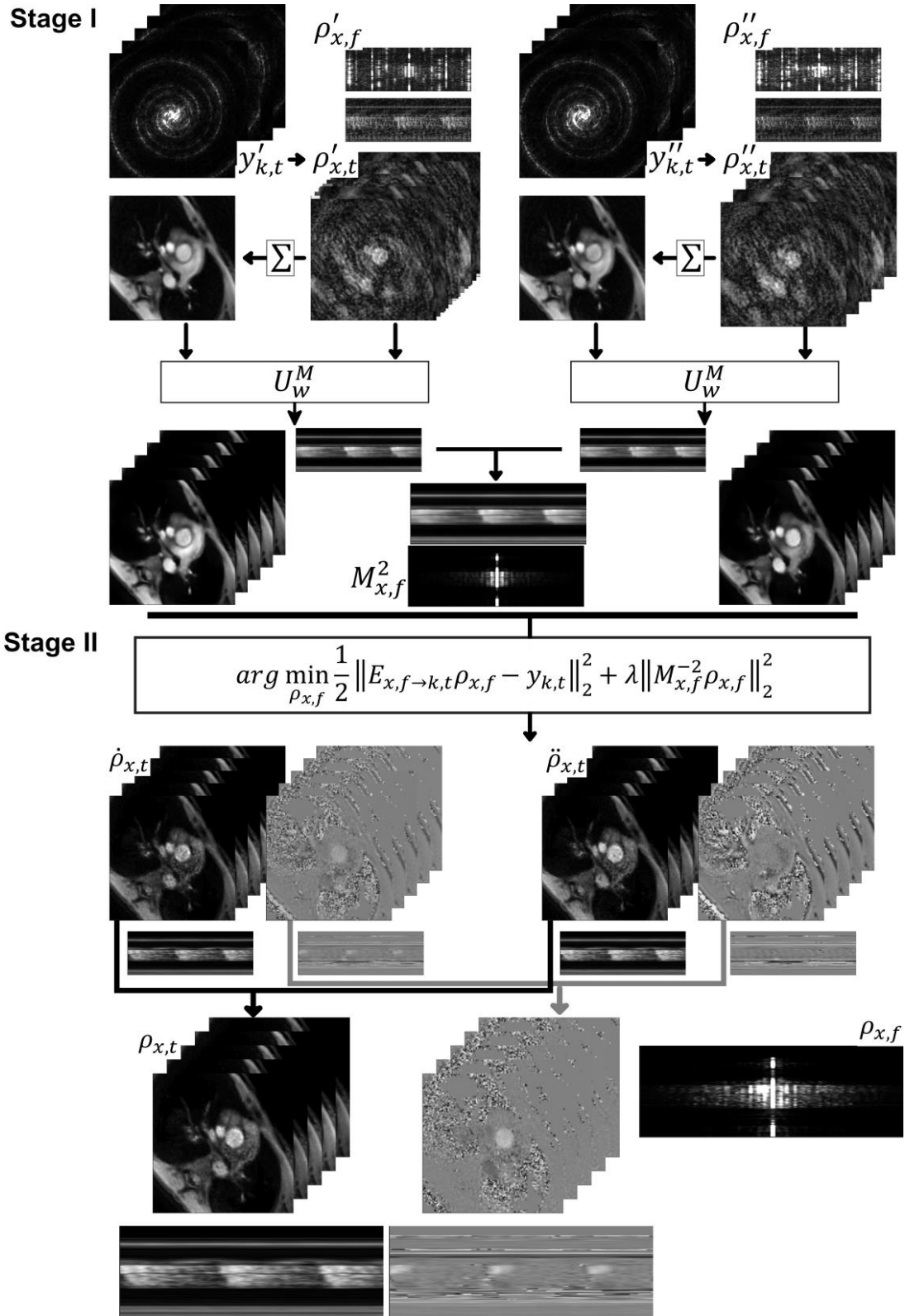

Fig. 2. The ML aided k-t SENSE processing.

Stage I – the prior ($M^2_{x,f}$) estimation using a CNN: both flow encoded ($y'_{k,t}$) and compensated ($y''_{k,t}$) data were processed as described in Eq. (2). The gridded multi-coil combined data ($\rho'_{x,f}$, $\rho''_{x,f}$) were time averaged ($\Sigma$) and the magnitudes were used as inputs for $U_w^M$. The $U_w^M$ magnitude predictions of the flow data were combined for the final x-f signal estimation. Stage II – k-t SENSE: the linear conjugate gradient solver was used to minimise [1] and produce the flow encoded and compensated complex-valued results ($\dot{\rho}_{x,t}, \ddot{\rho}_{x,t}$). These were then combined to produce the final PCMR results ($\rho_{x,t}$). The results x-f magnitude is presented ($\rho_{x,f}$).

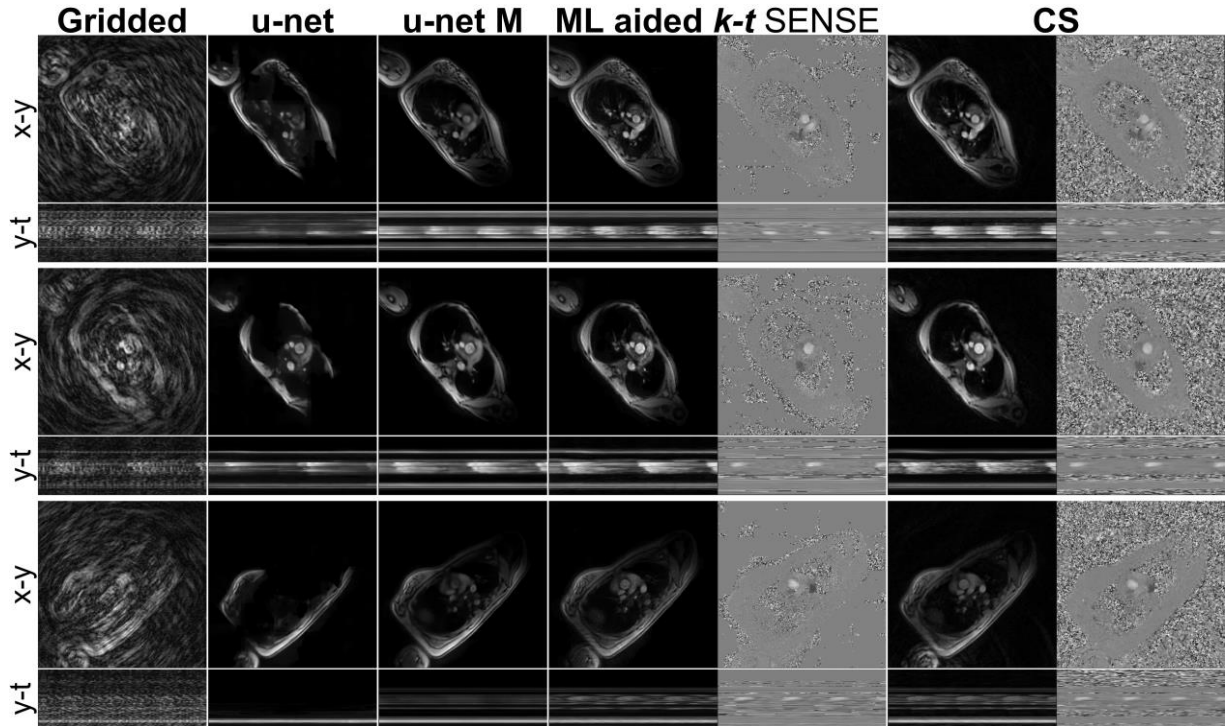

Fig. 3. Imaging results.

$U_w$ reconstructions presented with smaller or larger image artefacts: visible reconstruction patch boundary and signal removal. These are not visible on the $U_w^M$ results. In two cases $U_w$ removed heart structures (i.e. the bottom row). In these hard cases temporal blurring can be observed in the $U_w^M$ results. This had a small effect on the k-t SENSE magnitude results. However, it resulted in blurring of the extracted phase data Fig. 4 (2a-d).

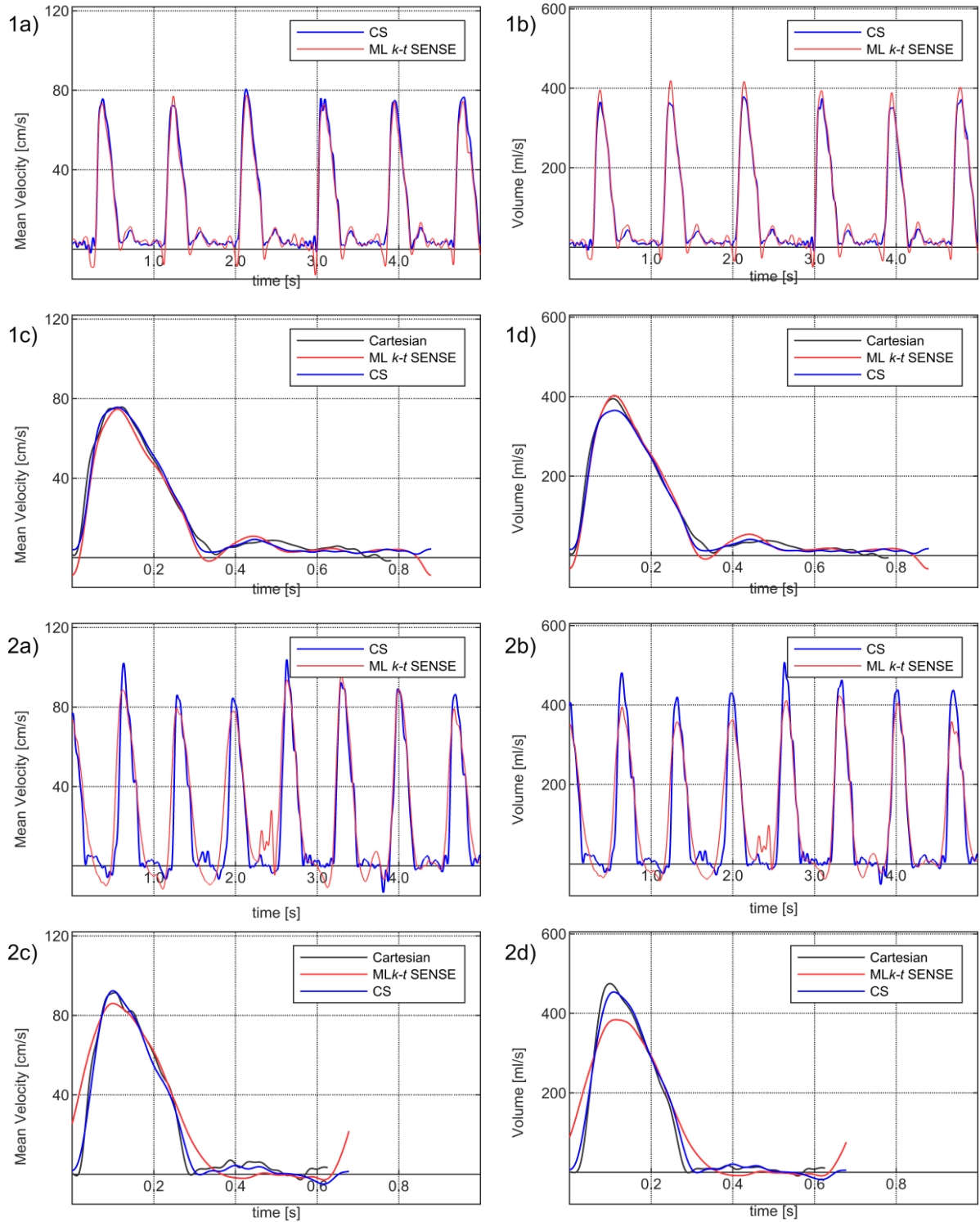

Fig. 4. The flow curves examples for two patients.
(1a-d) a comparison of mean velocity (a, c) and volume (b, d) curves extracted from the PCMR results. In general sharper slopes and peaks of the curves in the ML aided k-t SENSE results can be observed. (2a-d) one of the two low signal cases (Fig. 3 – bottom) that resulted in a sub-optimal $M_{x,f}^2$ prediction and blurring of the curves.

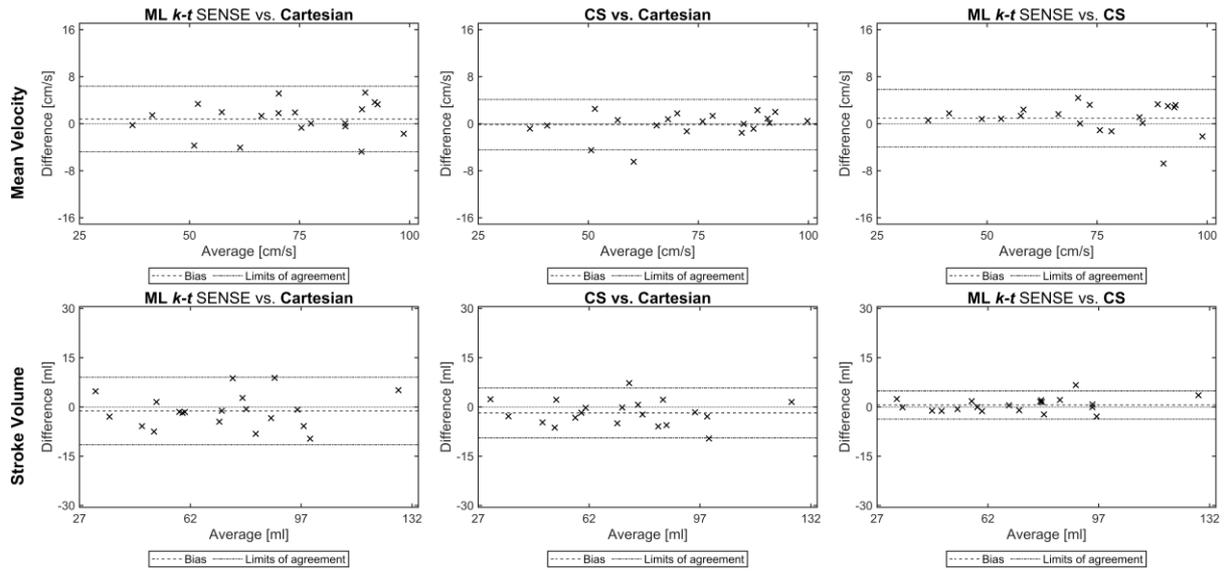

Fig. 5. Flow quantification results.

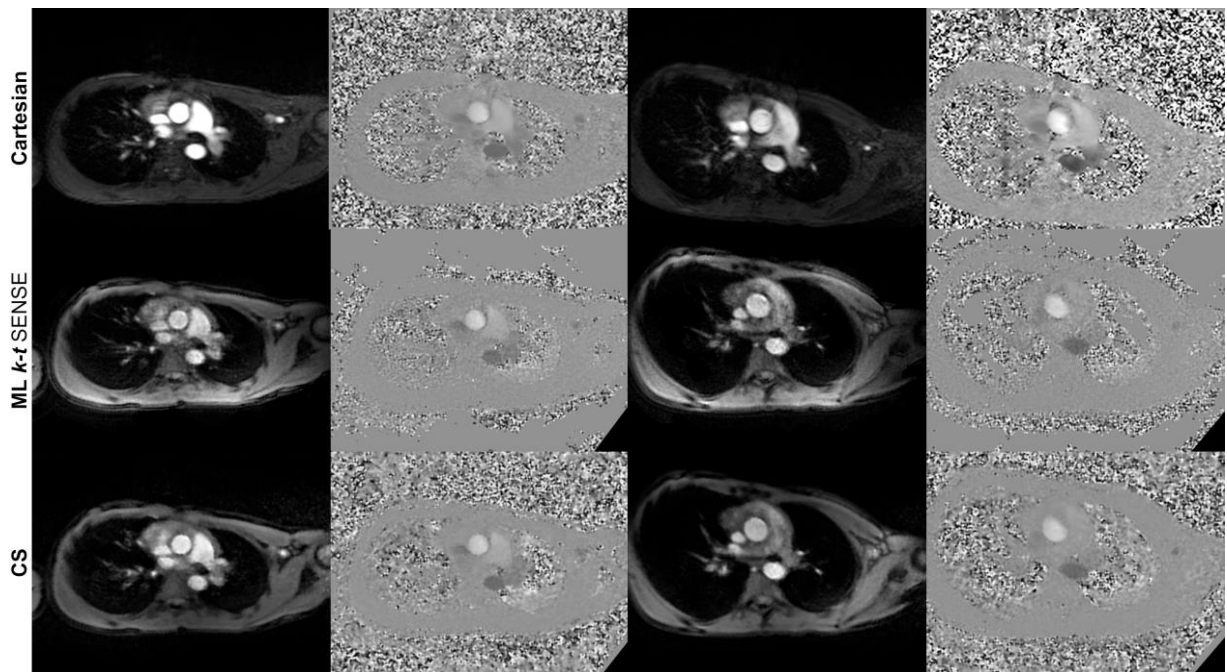

Fig. 6. The imaging results examples from two patients.

**Supporting Information**

Supporting Information Video S1 the Cartesian results

Supporting Information Video S2 the ML aided k-t SENSE results

Supporting Information Video S3 the Compressive Sensing results

Supporting Information Video S4 the Cartesian results

Supporting Information Video S5 the ML aided k-t SENSE results

Supporting Information Video S6 the Compressive Sensing results